\documentstyle[aps,multicol,prl,epsf]{revtex}
\title{Broadcasting of entanglement via local copying}
\author{V. Bu\v{z}ek$^{1,2}$,  V. Vedral$^{1}$, M. B. Plenio$^{1}$,
P.L.Knight$^{1}$, and
M. Hillery$^{3}$}
\address{
$^{1}$ Optics Section, The Blackett Laboratory,
Imperial College, London SW7 2BZ,  England\newline
$^{2}$ Institute of Physics, Slovak Academy of Sciences, Dubravsk\'a
cesta 9, 842 28 Bratislava, Slovakia\newline
$^{3}$Department
of Physics and Astronomy, Hunter College, CUNY,
695 Park Avenue, New York, NY 10021, USA
}
\date{January 20, 1997}

\begin{document}

\maketitle
\begin{abstract}
We show that inseparability of quantum states can be partially
broadcasted (copied, cloned)
with the help of {\em local} operations, i.e. distant parties
sharing
an entangled pair of spin 1/2 states can generate two pairs of
partially {\em nonlocally} entangled states  using only {\em local} 
operations.
This procedure can be viewed as an inversion
of quantum purification procedures.
\end{abstract}
\begin{multicols}{2}
\narrowtext
\section{INTRODUCTION}

The laws of quantum mechanics impose restrictions on manipulations with
quantum information.
These restrictions can on the one hand be fruitfully utilized in quantum
cryptography \cite{Ekert}. On the other hand they put limits on the
precision with which quantum-mechanical measurements or copying
(broadcasting, cloning) of quantum information can be performed
\cite{Wootters1,Barnum,Buzek1}. One of the most important aspects of
quantum-information processing is that information can be ``encoded''
in nonlocal correlations (entanglement) between two separated
particles. The more ``pure'' is the quantum entanglement, the more
``valuable'' is the
given two-particle state. This explains current interest in {\em purification}
procedures \cite{Bennett}
by means of which one can extract pure quantum entanglement
from a partially entangled state.  In other words, it is possible to
{\em compress} locally	an amount of quantum information.
This is implemented as follows: two ``distant''
parties share a number of partially
entangled pairs. They each then apply {\em local} operations
on their own particles and depending on the outcomes (which they 
are allowed to communicate
classically) they agree on further actions. By doing this they are
able to reduce the initial ensemble to a smaller one but whose pairs are
more entangled. This has important implications in the field of quantum
cryptography as it immediately implies an unconditional security of
communication at the quantum level.

Our main motivation for the present work comes from the fact that local
compression of quantum correlations is possible. We now ask the opposite:
can quantum correlations be `decompressed'? Namely, can two parties
acting locally start with a number of highly entangled pairs and
end up with a greater number of pairs with lower entanglement? This,
if possible, would also be of great operational value in determining
the amount of entanglement of a certain state \cite{Vedral}. For if
we could optimally `split' the original entanglement of a single
pair into two pairs equally entangled (e.g. having the same state) we have a
means of defining half the entanglement of the original pair.

We may view the process of decompression of quantum entanglement (i.e.,
inseparability) as a {\em local} copying (broadcasting, cloning) of
nonlocal quantum correlations. In this case
one might raise the question whether it is possible to
clone partially quantum entanglement using only {\em local} operations.
When we ask the question whether inseparability can be broadcast
via local copying  we mean
the following: Let two distant parties share an inseparable state
$\hat{\rho}_{a_{I}a_{II}}^{(id)}$. Now manipulate the two systems $a_{I}$
and $a_{II}$ {\em locally}, e.g. with the help of two distant quantum copiers
$X_I$ and $X_{II}$. These two quantum copiers are supposed
to be initially uncorrelated (or, more generally, they can be
classically correlated, i.e. the density operator $\hat{\rho}_{x_{I}x_{II}}$
describing the input state of two quantum copiers  is separable).
The quantum copier $X_{I}$ ($X_{II}$) copies the quantum subsystem
$a_{I}$ ($a_{II}$) such that at the output two systems $a_{I}$ and $b_{I}$
($a_{II}$ and $b_{II}$) are produced (see Fig. 1). As a result of this
copying we obtain out of
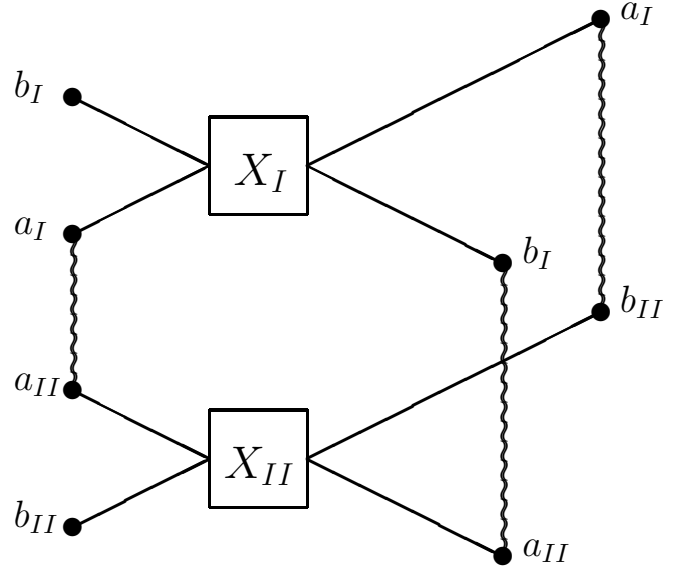
\begin{figure}[hbt]
\newcounter{cms}
\setlength{\unitlength}{1.3mm}
\begin{picture}(30,38)
\thicklines
\put(5,7){\circle*{2}}
\put(5,23){\circle*{2}}
\Large
\put(-1,7){\makebox(5,5)[bl]{$a_{II}$}}
\put(-1,23){\makebox(5,5)[bl]{$a_I$}}
\normalsize
\large
\multiput(4.6,21)(0,-2.){8}{\makebox(5,5)[bl]{$\wr$}}
\multiput(4.8,21)(0,-2.){8}{\makebox(5,5)[bl]{$\wr$}}
\put(5,7.05){\line(2,-1){14}}
\put(5,6.95){\line(2,-1){14}}
\put(5,23.05){\line(2,1){14}}
\put(5,22.95){\line(2,1){14}}
\put(5,-7){\circle*{2}}
\put(5,37){\circle*{2}}
\Large
\put(-1,-7){\makebox(5,5)[bl]{$b_{II}$}}
\put(-1,37){\makebox(5,5)[bl]{$b_I$}}
\normalsize
\put(5,-7.05){\line(2,1){14}}
\put(5,-6.95){\line(2,1){14}}
\put(5,37.05){\line(2,-1){14}}
\put(5,36.95){\line(2,-1){14}}
\LARGE
\put(19,-5){\line(0,1){10}}
\put(29,-5){\line(0,1){10}}
\put(19,5){\line(1,0){10}}
\put(19,-5){\line(1,0){10}}
\put(20.5,-2){\makebox(5,5)[bl]{$X_{II}$}}
\put(19,25){\line(0,1){10}}
\put(29,25){\line(0,1){10}}
\put(19,35){\line(1,0){10}}
\put(19,25){\line(1,0){10}}
\put(21.5,28){\makebox(5,5)[bl]{$X_{I}$}}
\normalsize
\put(29,29.95){\line(2,1){30}}
\put(29,30.05){\line(2,1){30}}
\put(29,29.95){\line(2,-1){20}}
\put(29,30.05){\line(2,-1){20}}
\put(29,-0.05){\line(2,1){30}}
\put(29,0.05){\line(2,1){30}}
\put(29,-0.05){\line(2,-1){20}}
\put(29,0.05){\line(2,-1){20}}
\put(59,45){\circle*{2}}
\put(49,20){\circle*{2}}
\Large
\put(61,45){\makebox(5,5)[bl]{$a_{I}$}}
\put(51,20){\makebox(5,5)[bl]{$b_{I}$}}
\normalsize
\put(59,15){\circle*{2}}
\put(49,-10){\circle*{2}}
\Large
\put(61,15){\makebox(5,5)[bl]{$b_{II}$}}
\put(51,-10){\makebox(5,5)[bl]{$a_{II}$}}
\normalsize
\large
\multiput(58.6,43)(0,-2.){15}{\makebox(5,5)[bl]{$\wr$}}
\multiput(58.8,43)(0,-2.){15}{\makebox(5,5)[bl]{$\wr$}}
\multiput(48.6,18)(0,-2.){15}{\makebox(5,5)[bl]{$\wr$}}
\multiput(48.8,18)(0,-2.){15}{\makebox(5,5)[bl]{$\wr$}}

\end{picture}\\[1.1cm]
\caption{\narrowtext An entangled pair of spin-1/2 particles $a_{I},a_{II}$
is shared
by two distant parties $I$ and $II$ which then perform local operations
using two quantum copiers $X_1$ and $X_2$. Each party obtains two
output particles which are in a separable state while the spatially
separated pairs $a_I,b_{II}$ and $a_{II},b_{I}$ are entangled.}
\end{figure}
\noindent
two systems $a_{I}$ and $a_{II}$ four systems described
by a density operator $\hat{\rho}_{a_{I}b_{I}a_{II}b_{II}}^{(out)}$.
If the states $\hat{\rho}_{a_{I}b_{II}}^{(out)}$ and
$\hat{\rho}_{a_{II}b_{I}}^{(out)}$ are {\em inseparable}
while the states $\hat{\rho}_{a_{I}b_{I}}^{(out)}$ and
$\hat{\rho}_{a_{II}b_{II}}^{(out)}$  which are	produced locally
are {\em separable}, then we say that we have partially broadcasted
(cloned, split)
the entanglement (inseparability) that was present in the input state.
As we said earlier, this broadcasting of inseparability can be viewed
as an inversion of the distillation protocol. The advantage of our operational
definition is that we impose the inseparability condition only between
two spins-1/2 (i.e., either on spins  $a_{I}$ and $b_{II}$, or
$a_{II}$ and $b_{I}$). Obviously, due to the quantum nature
of copying employed in our scheme, multiparticle quantum
correlations between {\em pairs} of spins $a_{I}b_{II}$ and
$a_{II}b_{I}$ (i.e., each of these  systems is described in 4-D Hilbert
space)	may appear at the output. But presently there do
not exist  strict criteria which  would allow to specify whether
these systems are inseparable (see below) and, consequently, it
would be impossible to introduce
operational definition of the inverse of the distillation protocol
based on multiparticle inseparability.

In this paper we
 show that the decompression of initial quantum entanglement
is indeed possible, i.e. that from a
pair of entangled particles  we can, by local
operations, obtain two less entangled pairs. Therefore entanglement can be
copied locally, i.e. the inseparability can be partially broadcasted.

\section{INSEPARABILITY AND PERES-HORODECKI THEOREM}

We first recall that
a density operator of two subsystems is
inseparable if it {\em cannot} be written as the convex sum
\begin{eqnarray}
	\hat{\rho}_{a_{I}a_{II}}  = \sum_m w^{(m)} \hat{\rho}_{a_{I}}^{(m)}
	\otimes \hat{\rho}_{a_{II}}^{(m)}.
	\label{8}
\end{eqnarray}
Inseparability is one of the most fundamental quantum phenomena, which,
in particular, may result in the violation of Bell's inequality (to be
specific, a separable system always satisfy Bell's inequality, but the
contrary is not necessarily true). Note that distant parties
cannot prepare an inseparable state from a separable state if they only
use local operations and classical communications.

We will not address the question of copying entanglement in its most
general form, but will rather focus our attention on copying of the 
entanglement
of spin-1/2 systems.
In this case, we can explicitly describe the transformations that are
necessary to broadcast entanglement . Moreover, in the case of two spins-1/2
we can effectively utilize the Peres-Horodecki theorem
\cite{Peres,Horodecki-sep} which states that the positivity of the
partial transposition of a state is {\em necessary} and {\em sufficient} for
its separability. Before we proceed further we briefly described how to
``use'' this theorem: The density matrix associated with the density operator
of two spins-1/2 can be written as
\begin{eqnarray}
	{\rho}_{m\mu,n\nu}  = \langle e_m|\langle f_\mu|\hat{\rho}
	| e_n\rangle|f_\nu\rangle,
\label{11}
\end{eqnarray}
where $\{ |e_m\rangle\}$ ($\{|f_{\mu}\rangle\}$) denotes an orthonormal basis
in the Hilbert space of the first (second) spin-1/2 (for instance,
$|e_0\rangle = |0\rangle_a$; $|e_1\rangle = |1\rangle_a$,  and
$|f_0\rangle = |0\rangle_b$; $|f_1\rangle = |1\rangle_b$). The
partial transposition  $\hat{\rho}^{T_2}$ of $\hat{\rho}$ is defined as
\begin{eqnarray}
	{\rho}_{m\mu,n\nu}^{T_2}  = {\rho}_{m\nu,n\mu}.
	\label{12}
\end{eqnarray}
Then the necessary and sufficient condition for the state $\hat{\rho}$
of two spins-1/2 to be inseparable is that at least one of the eigenvalues
of the partially transposed operator (\ref{12}) is negative. This
is equivalent to  the condition that at least one of  the two
determinants
\begin{eqnarray}
W_3 & = & \det\left(\begin{array}{ccc}
	\rho_{00,00}^{T_2} &  \rho_{00,01}^{T_2} & \rho_{00,10}^{T_2} \\
	\rho_{01,00}^{T_2} &  \rho_{01,01}^{T_2} & \rho_{01,10}^{T_2} \\
	\rho_{10,00}^{T_2} &  \rho_{10,01}^{T_2} & \rho_{10,10}^{T_2}
	\end{array}\right) \label{13a}\\[.25cm]
W_4 & = & \det\{\rho^{T_2}\}
\label{13b}
\end{eqnarray}
is negative. In principle one would also have to check the positivity of
the sub determinants $W_1=\rho_{00,00}^{T_2}$ and
$W_2=\rho_{00,00}^{T_2}\rho_{01,01}^{T_2} -\rho_{00,01}^{T_2}\rho_{01,00}^{T_2}$.
However, they are positive because the density operator $\hat{\rho}$ is
positive.
In this paper we deal exclusively with non-singular
operators $\rho^{T_2}$. Consequently, we do not face any problem which
may arise when	$\rho^{T_2}$ are singular.

\section{QUANTUM COPYING AND NO-BROADCASTING THEOREM}

In the realm of quantum physics
there does not exist a process which would allow us to copy (clone,
broadcast) an {\em arbitrary} state with perfect accuracy
\cite{Wootters1,Barnum,Buzek1}. What this means is that if the original
system is prepared in an arbitrary state $\hat{\rho}^{(id)}_a$, then
it is {\em impossible} to design a transformation
\begin{eqnarray}
\hat{\rho}^{(id)}_a \rightarrow  \hat{\rho}^{(out)}_{ab},
\label{1}
\end{eqnarray}
where $\hat{\rho}^{(out)}_{ab}$ is the density operator of the combined
original-copy quantum system after copying such that
\begin{eqnarray}
{\rm Tr}_b \hat{\rho}^{(out)}_{ab} =  \hat{\rho}^{(id)}_a ;\qquad
{\rm Tr}_a \hat{\rho}^{(out)}_{ab} =  \hat{\rho}^{(id)}_b.
\label{2}
\end{eqnarray}
This is the content of the {\em no-broadcasting} theorem which has been
recently proven by Barnum, Caves, Fuchs, Jozsa, and Schumacher
\cite{Barnum}. The stronger form of broadcasting, when
\begin{eqnarray}
\hat{\rho}^{(out)}_{ab}   = \hat{\rho}^{(id)}_{a} \otimes
\hat{\rho}^{(id)}_{b}
\label{3}
\end{eqnarray}
is denoted as the {\em cloning} of quantum states. Wootters and Zurek
\cite{Wootters1} were the first to point out that the cloning
of an {\em arbitrary} pure state is impossible. To be more specific,
the no-broadcasting and no-cloning theorems allow us to copy a single
{\em a priori} known state with absolute accuracy. In fact also two states
can be precisely copied if  it is {\em a priori} known that they are
orthogonal. But if no {\em a priori} information about the copied
(i.e., original) state is known, then precise copying (broadcasting)
is impossible.

Even though ideal copying is prohibited by the laws of quantum mechanics,
it is still possible to imagine quantum copiers which produce reasonably
good copies without destroying the original states too much. To be specific,
instead of imposing unrealistic constraints
on outputs of quantum copiers given by Eqs.(\ref{2}) and (\ref{3}), one
can adopt a more modest approach and give an operational definition
of a quantum copier. For instance, a reasonable quantum copier can be
specified by  three conditions:\newline
{\bf (i)} States of the original system and its quantum copy at the
output of the quantum copier, described by density operators
$\hat{\rho}^{(out)}_{a}$ and $\hat{\rho}^{(out)}_{b}$, respectively,
are identical, i.e.,
\begin{eqnarray}
\hat{\rho}^{(out)}_{a}	 = \hat{\rho}^{(out)}_{b}
\label{4}
\end{eqnarray}
{\bf (ii)} Once no {\em a priori} information about the {\em in}-state
of the original system is available, then it is reasonable to assume
that {\em all} pure states are copied equally well. One way to implement
this assumption is to design a quantum copier such that distances between
density operators of each system at the output ($\hat{\rho}^{(out)}_j$
where $j=a,b$)	and the ideal density operator $\hat{\rho}^{(id)}$
which describes the {\em in}-state of the original mode are input state
independent.  Quantitatively this means that if we employ the Bures
distance \cite{Bures}
\begin{eqnarray}
d_{B}(\hat{\rho}_{1};\hat{\rho}_{2}):=\sqrt{2}\left[1
- {\rm Tr}\left(\hat{\rho}^{1/2}_{1}\hat{\rho}_{2}\hat{\rho}^{1/2}_{1}
\right)^{1/2}\right]^{1/2}.
\label{5}
\end{eqnarray}
as a measure of distance between two operators, then the quantum copier
should be such that
\begin{eqnarray}
d_{B}(\hat{\rho}_{i}^{(out)};\hat{\rho}_{i}^{(id)})=
{\rm const.};\qquad i=a,b.
\label{6}
\end{eqnarray}
{\bf (iii)} It is important to note that the copiers we have in
mind are quantum devices. This means that even though we assume that
a quantum copier is initially disentangled (let us assume it is
in a pure state)  from the input system it is most likely that after
copying has been performed the copier will become entangled with the
output original+copy system. This entanglement is in part responsible
for an irreversible noise introduced into the output original+copy
system).  Consequently, $\hat{\rho}_{ab}^{(out)}\neq \hat{\rho}_{ab}^{(id)}$,
where $\hat{\rho}_{ab}^{(id)}=\hat{\rho}_{a}^{(id)} \otimes
\hat{\rho}_{b}^{(id)}$. Once again, if no {\em a priori} information
about the state $\hat{\rho}_{a}^{(id)}$ of the	input  system is
known, it is desirable to assume that the copier is such that the
Bures distance
between the actual output state $\hat{\rho}_{ab}^{(out)}$
of the original+copy system and the ideal output state 
$\hat{\rho}_{ab}^{(id)}$
is  input-state independent, i.e.
\begin{eqnarray}
	d_{B}(\hat{\rho}_{ab}^{(out)};\hat{\rho}_{ab}^{(id)})=
{\rm const.}
\label{7}
\end{eqnarray}

The copying process as specified by conditions (i)-(iii) can be understood
as  broadcasting in a weak sense, i.e., it is not perfect but it
can serve to some purpose when it is desirable to copy (at least
partially) quantum information without destroying it completely
(eavesdropping is one of the examples \cite{Fuchs1}).

The action of the quantum copier for spins-1/2 which satisfies the
conditions (i)-(iii)  can be
described in  terms of a unitary transformation of two basis
vectors $| 0\rangle_{a}$ and $| 1\rangle_{a}$ of the original system.
This transformation can be represented as \cite{Buzek1}
\begin{eqnarray}
\begin{array}{rcl}
| 0\rangle_{a}|0\rangle_{b}| Q\rangle_{x} & \longrightarrow &
\sqrt{\frac{2}{3}} | 00\rangle_{ab}|\uparrow\rangle_x
+\sqrt{\frac{1}{3}} | +\rangle_{ab}|\downarrow\rangle_{x};\\
| 1\rangle_{a}|0\rangle_{b}| Q\rangle_{x} & \longrightarrow &
\sqrt{\frac{2}{3}} | 11\rangle_{ab}|\downarrow\rangle_x
+\sqrt{\frac{1}{3}} | +\rangle_{ab}|\uparrow\rangle_x,
\end{array}
\label{14}
\end{eqnarray}
where $| Q\rangle_x$ describes the initial state of the quantum copier,
and $|\uparrow\rangle_x$ and $|\downarrow\rangle_x$ are two orthonormal
vectors in the Hilbert space of the quantum copier.
In Eq.(\ref{14}) we use the notation
such that $|e_me_n\rangle_{ab}=|e_m\rangle_{a}\otimes|e_n\rangle_{b}$
and $|+\rangle_{ab}=(|01\rangle_{ab}+|10\rangle_{ab})/\sqrt{2}$.
We do not specify the in state of the mode $b$	 in Eq.(\ref{14}).
In our discussion there is no need to specify this state. Obviously,
in real physical processes the in state of the mode $b$ may play an
important role.
In what follows, unless it may cause confusion,
we will omit subscripts indicating the subsystems.

\section{BROADCASTING OF INSEPARABILITY}

Now we present the basic operation necessary to copy entanglement
locally for spins-1/2. The scenario is as follows.
Two parties $X_I$ and $X_{II}$
share a pair of particles prepared in a state
\begin{eqnarray}
|\Psi\rangle_{a_I a_{II}} = \alpha |00\rangle_{a_I a_{II}}
+\beta |11\rangle_{a_I a_{II}},
	\label{4.1}
\end{eqnarray}
where we assume $\alpha$ and $\beta$ to be real and $\alpha^2+\beta^2=1$.
The state (\ref{4.1})	is inseparable for
all values of $\alpha^2$ such that $0<\alpha^2<1$, because one of the two
determinants $W_j$ from eqs.(\ref{13a}-\ref{13b}) is negative.
Now we assume that the system $a_I$ ($a_{II}$) is locally copied by
the quantum copier $X_I$ ($X_{II}$) operating according to the
transformations (\ref{14}). As the result of the copying we obtain a
composite system of four spins-1/2 described by the density operator
$\hat{\rho}_{a_{I}b_{I}a_{II}b_{II}}^{(out)}$. We are now interested to
see two properties of this output state. Firstly both the state
$\hat{\rho}_{a_{I}b_{II}}^{(out)}$ and $\hat{\rho}_{a_{II}b_{I}}^{(out)}$
should be inseparable simultaneously for at least some values of $\alpha$
and secondly the states $\hat{\rho}_{a_{I}b_{I}}^{(out)}$ and
$\hat{\rho}_{a_{II}b_{II}}^{(out)}$ should be separable simultaneously
for some values of $\alpha$ for which
$\hat{\rho}_{a_{I}b_{II}}^{(out)}$ and $\hat{\rho}_{a_{II}b_{I}}^{(out)}$
are inseparable.

Using the transformation (\ref{14}) we find the local output of the quantum
copier $X_I$ to be described by the density operator
\begin{eqnarray}
	\hat{\rho}^{(out)}_{a_Ib_{I}}= \frac{2\alpha^2}{3}|00\rangle\langle 00|
	+\frac{1}{3}|+\rangle\langle +|
	+\frac{2\beta^2}{3} |11\rangle\langle 11|,
	\label{23}
\end{eqnarray}
while the nonlocal pair of output particles is in the state described by
the density operator
\begin{eqnarray}
	\hat{\rho}^{(out)}_{a_{I}b_{II}} &=&
	\frac{24\alpha^2+1}{36} |00\rangle\langle 00|
	+\frac{24\beta^2+1}{36} |11\rangle\langle 11|
	\label{28}\\
&&\hspace*{-1.cm} + \frac{5}{36}(|01\rangle\langle 01|+|10\rangle\langle 10|)
	+ \frac{4\alpha\beta}{9}(|00\rangle\langle 11|+|11\rangle\langle 00|).
	\nonumber
\end{eqnarray}
We note that due to the symmetry between the systems $I$ and $II$ we have that
$\hat{\rho}^{(out)}_{a_{I}b_{I}}=\hat{\rho}^{(out)}_{a_{II}b_{II}}$
and
$\hat{\rho}^{(out)}_{a_{I}b_{II}}=\hat{\rho}^{(out)}_{a_{II}b_{I}}$.

Now we check for which values of $\alpha$ the density operator
$\hat{\rho}^{(out)}_{a_{I}b_{II}}$ is inseparable.
>From the determinants in Eqs.(\ref{13a}-\ref{13b}) associated with this
density operator it immediately follows
that $\hat{\rho}^{(out)}_{a_{I}b_{II}}$
is inseparable if
\begin{eqnarray}
	\frac{1}{2} - \frac{\sqrt{39}}{16} \leq \alpha^2 \leq
	\frac{1}{2} + \frac{\sqrt{39}}{16}\;\; .
	\label{30}
\end{eqnarray}
On the other hand from Eq.(\ref{23})
we find that $\hat{\rho}_{a_{I}b_{I}}^{(out)}$
is separable if
\begin{eqnarray}
	\frac{1}{2} - \frac{\sqrt{48}}{16} \leq \alpha^2 \leq
	\frac{1}{2} + \frac{\sqrt{48}}{16}\;\; .
	\label{31}
\end{eqnarray}
Comparing eqs. (\ref{30}) and (\ref{31}) we observe that 
$\hat{\rho}_{a_{I}b_{I}}^{(out)}$ is {\em separable} if
$\hat{\rho}_{a_{I}b_{II}}^{(out)}$ is {\em inseparable}. This finally
proves that it is possible to clone partially quantum entanglement using
only local operations and classical communication. Note that any other initial
state obtained by applying local unitary transformation will yield
the same result.

This last result clearly illustrates the fact that for given values
of $\alpha^2$  the inseparability of the input state can be
broadcasted by performing local operations. To appreciate
more clearly this result we turn our attention to the copying of
a separable state of the form
\begin{eqnarray}
\hat{\rho}^{(in)}_{a_{I}a_{II}}=\sum_i \, w_i \, \hat{\rho}^{(in)}_{a_{I,i}}
\otimes\hat{\rho}^{(in)}_{a_{II,i}},
\label{26}
\end{eqnarray}
In this case it is easily seen that the output of our procedure is
of the form
\begin{eqnarray}
\hat{\rho}^{(out)}_{a_{I}b_{I}a_{II}b_{II}}=\sum_i \, u_i \,
\hat{\rho}^{(out)}_{a_{I,i}b_{I,i}}
\otimes\hat{\rho}^{(out)}_{a_{II,i}b_{II,i}},
\label{27}
\end{eqnarray}
from which it follows that in this case the output
$\hat{\rho}^{(out)}_{a_{I}b_{II}}$ is always separable, i.e.,
\begin{eqnarray}
\hat{\rho}^{(out)}_{a_{I}b_{II}}=\sum_i \,  v_i \, 
\hat{\rho}^{(out)}_{a_{I,i}}
\otimes\hat{\rho}^{(out)}_{b_{II,i}}.
\label{32}
\end{eqnarray}
This illustrates the fact that the inseparability cannot be produced
by two distant parties operating locally and
who can communicate only classically. This result is not only related to
our procedure but is easily seen to be valid for general local operations and
classical communications.

\section{CONCLUSIONS}

In conclusion, using a simple set of local operations which can be expressed
in terms of quantum state copying \cite{Buzek1} we have shown that
inseparability of quantum states can be {\em locally} copied with
the help of {\em local} quantum copiers. We will investigate elsewhere
how close the distilled copied states
$\hat{\tilde{\rho}}^{(out)}_{a_{I}b_{II}}$ and
$\hat{\tilde{\rho}}^{(out)}_{b_{I}a_{II}}$ are to the distilled input state
$\hat{\tilde{\rho}}^{(in)}_{a_{I}a_{II}}$ and in particular whether
the efficiency of the quantum copying can be improved when we do not
average over all possible output states of the quantum copier but perform
measurements on the quantum copier (conditional output states).
This will give us a qualitative measure how well a pure quantum entanglement
can be broadcasted. More importantly, we would like
to generalize our procedure such that any amount of initial entanglement,
no matter how small, can be split into two even less entangled states. We
now know that an equivalent of such a general procedure exists for
purification procedures \cite{Horodecki-dis}. This, when found, would
give us operational means of quantifying the amount of entanglement
\cite{Vedral}.

\vspace{1.5truecm}

{\bf Acknowledgements}\newline
This work was supported  by the United Kingdom Engineering
and Physical Sciences Research Council, by the grant agency VEGA
of the Slovak Academy of Sciences (under the project
2/1152/96), by the National Science
Foundation under the grant  INT 9221716, the European Union,
the Alexander von Humboldt Foundation and the Knight Trust.

\section*{APPENDIX.  }
In this paper we have utilized one nontrivial quantum-copier transformation
(\ref{14})
with the help of which broadcasting of entanglement via local copying
can be performed. Here we present a scheme by means of which one can
in principle determine
a class of local quantum-copier transformations such that local
outputs of quantum copiers are described by separable density operators
$\hat{\rho}^{(out)}_{a_{I}b_{I}}$ and $\hat{\rho}^{(out)}_{a_{II}b_{II}}$
while the nonlocal states $\hat{\rho}^{(out)}_{a_{I}b_{II}}$
and $\hat{\rho}^{(out)}_{a_{II}b_{I}}$ are inseparable.

The most general quantum-copier transformation for a single spin-1/2
has the form
\begin{eqnarray}
\begin{array}{rcl}
| 0\rangle_{a}| Q\rangle_{x} & \longrightarrow &
\sum_{i=1}^4 |R_i\rangle_{ab}|X_i\rangle_x;\\
| 1\rangle_{a}| Q\rangle_{x} & \longrightarrow &
\sum_{i=1}^4 |R_i\rangle_{ab}|Y_i\rangle_x,
\end{array}
\eqnum{A.1}
\label{A.1}
\end{eqnarray}
where $|R_i\rangle_{ab}$ ($i=1,...,4$) are four basis  vectors
in the four-dimensional Hilbert space of the output modes $a$ and $b$.
These vectors are defined as: $|R_1\rangle = |00\rangle$ ;
$|R_2\rangle = |01\rangle$; $|R_3\rangle = |10\rangle$,
and $|R_4\rangle = |11\rangle$. The output states $|X_i\rangle_x$ and
$|Y_i\rangle_x$ of the quantum copier in the basis of four orthonormal
quantum-copier states $|Z_i\rangle_x$ read:
\begin{eqnarray}
\begin{array}{rcl}
| X_i\rangle_{x} & = &
\sum_{k=1}^4 C^{(i)}_k|Z_k\rangle_x;\\
| Y_i\rangle_{x} & = &
\sum_{k=1}^4 D^{(i)}_k|Z_k\rangle_x.
\end{array}
\eqnum{A.2}
\label{A.2}
\end{eqnarray}
The amplitudes $C^{(i)}_k$ and $D^{(i)}_k$ specify the action of the
quantum copier under consideration. From the unitarity	of the transformation
(\ref{A.1}) three conditions on these amplitudes  follow:
\begin{eqnarray}
\begin{array}{rcl}
\sum_{k=1}^4 |C^{(i)}_k|^2 & = & 1;\\
\sum_{k=1}^4 |D^{(i)}_k|^2 & = & 1;\\
\sum_{k=1}^4 C^{(i)}_k D^{(i)}_k & = & 1.
\end{array}
\eqnum{A.3}
\label{A.3}
\end{eqnarray}

The further specification of the amplitudes $C^{(i)}_k$ and $D^{(i)}_k$
depends on the tasks which should be performed by the quantum copier
under consideration. This means that we have to specify these amplitudes
in terms of constraints imposed on
the output of the  copier. These constraints (which can take form of
specific equalities or inequalities) then define domains of
acceptable values of $C^{(i)}_k$ and $D^{(i)}_k$.

To be specific, let us assume that the entangled state (\ref{4.1}) is going
to be broadcasted by two {\em identical} local quantum copiers
defined by Eq.(\ref{A.1}). In this case the density operator
$\hat{\rho}^{(out)}_{a_{I}b_{I}a_{II}b_{II}}$
describing the four particle output of the two copiers reads
(in what follows we assume the amplitudes $C^{(i)}_k$ and $D^{(i)}_k$
to be real):
\begin{eqnarray}
\begin{array}{c}
\hat{\rho}^{(out)}_{a_{I}b_{I}a_{II}b_{II}} =
\sum_{i_{I}i_{II}j_{I}j_{II}} \sum_{k_{I}k_{II}}
 \omega_{k_{I}k_{II}}^{(i_{I}i_{II})}
\omega_{k_{I}k_{II}}^{(j_{I}j_{II})}\\
\mbox{~~~}\\
\times |R_{i_{I}}\rangle_{a_{I}b_{I}}\langle R_{j_{I}}| \,
|R_{i_{II}}\rangle_{a_{II}b_{II}}\langle R_{j_{II}}| ,
\end{array}
\eqnum{A.4}
\label{A.4}
\end{eqnarray}
where
\begin{eqnarray}
\omega_{kl}^{(ij)}=
\alpha C_{k}^{(i)} C_l^{(j)} + \beta D_k^{(i)} D_l^{(j)}.
\eqnum{A.5}
\label{A.5}
\end{eqnarray}
The local output of the quantum copier $X_I$ is now described by the
density operator $\hat{\rho}^{(out)}_{a_{I}b_{I}}$ which can be expressed
as
\begin{eqnarray}
\hat{\rho}^{(out)}_{a_{I}b_{I}}=
\sum_{i_{I}j_{I}}
 \Xi^{(i_{I}j_{I})}
|R_{i_{I}}\rangle_{a_{I}b_{I}}\langle R_{j_{I}}|,
\eqnum{A.6}
\label{A.6}
\end{eqnarray}
where the matrix elements $\Xi^{(i_{I}j_{I})}$ of this density
operator in the basis $|R_{i_{I}}\rangle_{a_{I}b_{I}}$	read
\begin{eqnarray}
 \Xi^{(i_{I}j_{I})}=
\sum_{k}
\alpha^2 C_{k}^{(i_I)} C_k^{(j_I)} + \beta^2 D_k^{(i_I)} D_k^{(j_I)}.
\eqnum{A.7}
\label{A.7}
\end{eqnarray}
In our discussion  of broadcasting of entanglement we have assumed
that local outputs  of quantum copiers $X_I$ and $X_{II}$ are
separable. This implies restrictions on the density operator
$\hat{\rho}^{(out)}_{a_{I}b_{I}}$,  i.e., the four eigenvalues of the
partially transposed operator $[\hat{\rho}^{(out)}_{a_{I}b_{I}}]^{T_2}$
have to be positive \cite{Peres,Horodecki-sep}. So these are four additional
constraints on the amplitudes  $C_{k}^{(i)}$ and $D_{k}^{(i)}$
[the first three constraints are given by Eq.(\ref{A.2})].
Further constraints are to be obtained from the assumption that the
density operator $\hat{\rho}^{(out)}_{a_{I}b_{II}}$
is inseparable. The explicit expression for this density operator
can be expressed in the form
\begin{eqnarray}
\hat{\rho}^{(out)}_{a_{I}b_{II}}=
\sum_{i_{I}j_{II}}
 \Omega^{(i_{I}j_{II})}
|R_{i_{I}}\rangle_{a_{I}b_{II}}\langle R_{j_{II}}|,
\eqnum{A.8}
\label{A.8}
\end{eqnarray}
where the diagonal matrix elements $\Omega^{(i_{I}j_{II})}$
read:
\begin{eqnarray}
\begin{array}{rl}
\Omega^{(1,1)} = \sum_{kl} & [\omega^{(1,1)}_{kl}\omega^{(1,1)}_{kl}
			+ \omega^{(2,1)}_{kl}\omega^{(2,1)}_{kl}\\
	      & 	+ \omega^{(1,3)}_{kl}\omega^{(1,3)}_{kl}
			+ \omega^{(2,3)}_{kl}\omega^{(2,3)}_{kl}] \\
\Omega^{(2,2)} = \sum_{kl} &[\omega^{(1,2)}_{kl}\omega^{(1,2)}_{kl}
			+ \omega^{(2,2)}_{kl}\omega^{(2,2)}_{kl}\\
	      & 	+ \omega^{(1,4)}_{kl}\omega^{(1,4)}_{kl}
			+ \omega^{(2,4)}_{kl}\omega^{(2,4)}_{kl}]\\
\Omega^{(3,3)} = \sum_{kl} & [\omega^{(3,1)}_{kl}\omega^{(3,1)}_{kl}
			+ \omega^{(4,1)}_{kl}\omega^{(4,1)}_{kl}\\
	      & 	+ \omega^{(3,3)}_{kl}\omega^{(3,3)}_{kl}
			+ \omega^{(4,3)}_{kl}\omega^{(4,3)}_{kl}]\\
\Omega^{(4,4)} = \sum_{kl} & [\omega^{(4,2)}_{kl}\omega^{(4,2)}_{kl}
			+ \omega^{(3,2)}_{kl}\omega^{(3,2)}_{kl}\\
	      & 	+ \omega^{(3,4)}_{kl}\omega^{(3,4)}_{kl}
			+ \omega^{(4,4)}_{kl}\omega^{(4,4)}_{kl} ]\\
\end{array}
\eqnum{A.9}
\label{A.9}
\end{eqnarray}
For the off-diagonal matrix elements we find
\begin{eqnarray}
\begin{array}{rl}
\Omega^{(2,1)}	 & = \sum_{kl}	[\omega^{(2,2)}_{kl}\omega^{(2,1)}_{kl}
			+ \omega^{(1,2)}_{kl}\omega^{(1,1)}_{kl}\\
	      & 	+ \omega^{(1,4)}_{kl}\omega^{(1,3)}_{kl}
		    + \omega^{(2,4)}_{kl}\omega^{(2,3)}_{kl}] = \Omega^{(1,2)} 
\\
\Omega^{(3,1)}	 & = \sum_{kl}	[\omega^{(3,1)}_{kl}\omega^{(1,1)}_{kl}
			+ \omega^{(4,1)}_{kl}\omega^{(2,1)}_{kl}\\
	      & 	+ \omega^{(3,3)}_{kl}\omega^{ 1,3}_{kl}
		    + \omega^{(4,3)}_{kl}\omega^{(2,3)}_{kl}] = \Omega^{(1,3)} 
\\
\Omega^{(4,1)}	 & = \sum_{kl}	[\omega^{(3,2)}_{kl}\omega^{(1,1)}_{kl}
			+ \omega^{(4,2)}_{kl}\omega^{(2,1)}_{kl}\\
	      & 	+ \omega^{(3,4)}_{kl}\omega^{(1,3)}_{kl}
		    + \omega^{(4,4)}_{kl}\omega^{(2,3)}_{kl}] = \Omega^{(1,4)} 
\\
\Omega^{(3,2)}	 & = \sum_{kl}	[\omega^{(1,2)}_{kl}\omega^{(3,1)}_{kl}
			+ \omega^{(2,2)}_{kl}\omega^{(4,1)}_{kl}\\
	      & 	+ \omega^{(1,4)}_{kl}\omega^{(3,3)}_{kl}
		    + \omega^{(2,4)}_{kl}\omega^{(4,3)}_{kl}] = \Omega^{(2,3)} 
\\
\Omega^{(4,2)}	 & = \sum_{kl}	[\omega^{(1,2)}_{kl}\omega^{(3,1)}_{kl}
			+ \omega^{(2,2)}_{kl}\omega^{(4,2)}_{kl}\\
	      & 	+ \omega^{(1,4)}_{kl}\omega^{(3,4)}_{kl}
		    + \omega^{(2,4)}_{kl}\omega^{(4,4)}_{kl}] = \Omega^{(2,4)} 
\\
\Omega^{(3,4)}	 & = \sum_{kl}	[\omega^{(3,1)}_{kl}\omega^{(3,2)}_{kl}
			+ \omega^{(4,1)}_{kl}\omega^{(4,2)}_{kl}\\
	      & 	+ \omega^{(3,3)}_{kl}\omega^{(3,4)}_{kl}
		    + \omega^{(4,3)}_{kl}\omega^{(4,4)}_{kl}] = \Omega^{(4,3)} 
\\
\end{array}
\eqnum{A.10}
\label{A.10}
\end{eqnarray}
If the density operator is supposed to be inseparable then at least one
of the eigenvalues of the partially transposed operator
$\hat{\rho}^{(out)}_{a_{I}b_{II}}$  has to be negative. This represents
another condition which specifies the amplitudes $C_k^{(i)}$
and $D_k^{(i)}$.

We have to note that the conditions we have derived result in
a set of nonlinear equations which are very difficult to solve explicitly.
Moreover, these equations do not specify the amplitudes uniquely,
so more constraints have to be found. Obviously, it will then become more
difficult to check whether there exist some amplitudes $C_k^{(i)}$
and $D_k^{(i)}$ which fulfill these constraints.

\end{multicols}

\end{document}